\documentclass[utf8]{frontiersFPHY} 

\setcitestyle{square} 
\usepackage{url,hyperref,lineno,microtype,subcaption}
\usepackage[onehalfspacing]{setspace}

\def\firstAuthorLast{Verscharen {et~al.}} 
\def\Authors{Verscharen, D.\,$^{1,2*}$, Wicks, R.~T.\,$^{3}$, Branduardi-Raymont, G.\,$^{1}$, Erd\'elyi, R.\,$^{4,5,6}$, Frontera, F.\,$^{7}$, G\"otz, C.\,$^{8}$, Guidorzi, C.\,$^{7}$, Lebouteiller, V.\,$^{9,10}$, Matthews, S.~A.\,$^{1}$, Nicastro, F.\,$^{11}$, Rae, I.~J.\,$^{3}$, Retin\`o, A.\,$^{12}$, Simionescu, A.\,$^{13,14,15}$, Soffitta, P.\,$^{16}$, Uttley, P.\,$^{17}$ and Wimmer-Schweingruber, R.~F.\,$^{18,19}$}
\def\Address{$^{1}$Mullard Space Science Laboratory, University College London, Dorking, UK\\
$^{2}$Space Science Center, University of New Hampshire, Durham, NH, USA\\
$^{3}$Department of Mathematics, Physics and Electrical Engineering, Northumbria University, Newcastle upon Tyne, UK\\
$^{4}$Solar Physics \& Space Plasma Research Centre, University of Sheffield, Sheffield, UK\\
$^{5}$Department of Astronomy, E\"otv\"os Lor\'and University, Budapest, Hungary\\
$^{6}$Gyula Bay Zolt\'an Solar Observatory (GSO), Hungarian Solar Physics Foundation (HSPF), Gyula, Hungary\\
$^{7}$Department of Physics and Earth Sciences, University of Ferrara, Ferrara, Italy\\
$^{8}$ESTEC, European Space Agency, Noordwijk, The Netherlands\\
$^{9}$AIM, CEA, CNRS, Universit\'e Paris-Saclay, Universit\'e Paris Diderot, Sorbonne Paris Cit\'e, Gif-sur-Yvette, France\\
$^{10}$Department of Physics and Astronomy, University of North Carolina, Chapel Hill, NC, USA\\
$^{11}$Italian National Institute for Astrophysics (INAF), Rome Astronomical Observatory, Rome, Italy\\
$^{12}$Laboratoire de Physique des Plasmas, \'Ecole Polytechnique, Palaiseau, France\\
$^{13}$SRON Netherlands Institute for Space Research, Utrecht, The Netherlands\\
$^{14}$  Leiden Observatory, Leiden University, Leiden, The Netherlands \\ 
$^{15}$  Kavli Institute for the Physics and Mathematics of the Universe (WPI), The University of Tokyo, Kashiwa, Japan\\
$^{16}$Italian National Institute for Astrophysics (INAF), Istituto di Astrofisica e Planetologia Spaziali, Rome, Italy\\
$^{17}$Anton Pannekoek Institute, University of Amsterdam, Amsterdam, The Netherlands\\
$^{18}$Institute of Experimental and Applied Physics, Kiel University, Kiel, Germany\\
$^{19}$National Space Science Center, Chinese Academy of Sciences, Beijing, China
}

\def\corrAuthor{Daniel Verscharen, Holmbury House, Holmbury St. Mary, Dorking, RH5~6NT, UK}

\def\corrEmail{d.verscharen@ucl.ac.uk}

\begin{document}
\onecolumn
\firstpage{1}

\title[The Plasma Universe]{The Plasma Universe: A Coherent Science Theme for Voyage 2050} 

\author[\firstAuthorLast ]{\Authors} 
\address{} 
\correspondance{} 

\extraAuth{}

\maketitle

In review of the White Papers from the Voyage 2050 process\footnote{All Voyage 2050 White Papers are available online at \href{https://www.cosmos.esa.int/web/voyage-2050/white-papers}{\url{https://www.cosmos.esa.int/web/voyage-2050/white-papers}}.} and after the public presentation of a number of these papers in October 2019 in Madrid, we as White Paper lead authors have identified a coherent science theme that transcends the divisions around which the Topical Teams are structured. This note aims to highlight this synergistic science theme and to make the Topical Teams and the Voyage 2050 Senior Committee aware of the wide importance of these topics and the broad support that they have across the worldwide science community.

Baryonic matter in the Universe is almost exclusively in the plasma state. It forms structures on a huge range of scales, reaching from the kinetic electron and ion microscales to the size of the entire observable Universe. These plasmas include very diverse objects such as magnetic cavities around comets, planetary magnetospheres, the solar atmosphere, the outer heliosphere, accretion discs around compact objects, galaxy-scale ``Fermi bubbles'', the intracluster medium, and the intergalactic medium permeating the cosmic web. The key difficulty in understanding of all these objects lies in the two-way nature of the intrinsic multi-scale physics of plasmas: processes on the largest scales affect the small-scale physics, and processes on the smallest scales affect the large-scale evolution of plasmas.

These multi-scale processes are united by \emph{fundamental physics questions} that underpin the physics addressed in all of the 18 White Papers referenced below, e.g.:
\begin{itemize}
\item How are electrons and ions heated and accelerated, and how is energy partitioned?
\item What is the role of the magnetic field?
\item What are the properties and roles of different energisation regions in plasma structures?
\item What is the role of plasma physics in the formation and evolution of different processes and objects including flux tubes, turbulence, waves, flows, jets, discs, magnetospheres, coronae, and halos? 
\item What are the effects of rapid and discontinuous processes such as shocks and reconnection?
\end{itemize}
The answers to these fundamental questions are very important for a wide range of \emph{processes in the Universe} including:
\begin{itemize}
\item accretion of matter onto compact objects,
\item cosmic-ray acceleration,
\item galaxy formation,
\item heat and energy transfer, conduction, diffusion, and turbulence in plasma flows on all scales, in intergalactic, interstellar, and interplanetary media,
\item magnetic-field generation through dynamo processes,
\item magnetospheric dynamics,
\item stellar activity and coronal dynamics, and
\item space weather.
\end{itemize}

We have specifically identified four fields of study in the proposed Voyage 2050 White Papers that are linked by this common theme:

\emph{Astronomy from the UV to soft and hard X-ray wavelengths} is a powerful tool to explore different parameter regimes and examples of plasma environments on large scales based on a whole-system overview. They allow us to identify plasma shocks, thermal processes in accretion flows onto compact objects such as neutron stars and black holes, the large-scale geometry of matter, and even elemental and charge-state composition through the effective use of spectroscopy and polarimetry  \citep{guidorzi19,frontera19,lebouteiller19,nicastro19,simionescu19,soffitta19,uttley19}.

\emph{Solar physics} investigates processes on intermediate scales and links the physics explored by X-ray and UV astronomy to the local environment of the solar system. It allows us to obtain detailed spectroscopic imagery of plasma phenomena that we can interpret directly  \citep{branduardi19,matthews19,peter19,erdelyi19}.

\emph{Heliospheric, magnetospheric, and cometary physics studies} of in-situ plasma phenomena such as the acceleration and heating of particles can be directly linked to larger structures with a good level of system-wide imagery and context \citep{branduardi19,matthews19,peter19,erdelyi19,goetz19,rae19,wimmer19,roussos19,mccrea19}.

\emph{In-situ plasma physics} explores the near-Earth plasma environment (e.g., pristine and shocked solar wind, bow shock, and magnetosphere) and the plasma environment around other solar-system objects. It allows us to analyse the detailed fundamental interactions and the micro-scale processes that determine the large-scale evolution and thermodynamics of matter \citep{branduardi19,goetz19,rae19,wimmer19,retino19,verscharen19}.

Although these science topics appear quite diverse and each White Paper will be evaluated on its own merit by their respective Topical Team, we emphasise that all of them will mutually benefit from each other. For instance, the interpretation of X-ray and UV observations, reaching from compact objects to the largest structures in the Universe, depends on a solid understanding of fundamental in-situ plasma physics. On the other hand, the in-situ plasma community will benefit from cross-disciplinary collaboration with plasma astrophysicists by studying a much wider range of plasma conditions, some of which cannot be studied in situ. The same benefit applies likewise to the solar, heliospheric, magnetospheric, and cometary fields. Moreover, numerical modelling of plasmas in different regimes with shared physical understanding will underpin much of the developments in these fields.

The synopsis above and the related Voyage 2050 White Papers show that a common and coherent science theme has emerged from the Voyage 2050 process. This theme is linked by the common interest across large parts of the ESA-science community in exploring structures in the Universe that are shaped by plasma processes across a large variety of scales. This science theme spans across all of the installed Topical Teams. We are convinced that the adoption of this coherent science theme by ESA through a programme of missions addressing plasma physics in its many forms will make transformative advances in our knowledge of fundamental plasma physics questions and of a wide range of processes that are of greatest importance for our understanding of the Universe.

\section*{Conflict of Interest Statement}
The authors declare that the research was conducted in the absence of any commercial or financial relationships that could be construed as a potential conflict of interest.

\section*{Author Contributions}
All authors contributed to the writing of this article.

\section*{Funding}
D.~V.~is supported by Science and Technology Facilities Council (STFC) Ernest Rutherford Fellowship ST/P003826/1 and STFC Consolidated Grant ST/S000240/1. R.~T.~W. and I.~J.~R. are supported by STFC Consolidated Grant ST/V006320/1. R.~E. is grateful to STFC (grant number ST/M000826/1) and the Royal Society for enabling this research. R.~E. also acknowledges the support received by the CAS President's International Fellowship Initiative Grant No.~2019VMA052 and the warm hospitality received at USTC of CAS, Hefei, where part of his contribution was made.

\section*{Acknowledgments}
Apart from minor edits, this article was submitted as a supporting statement in response to the European Space Agency's (ESA's) long-term planning cycle Voyage 2050.  We are grateful to ESA's Directorate of Science, the Science Programme Committee (SPC), the Voyage 2050 Senior Committee, and the Voyage 2050 Topical Teams for the consideration of the community's input. 

\bibliographystyle{frontiersinHLTH&FPHY} 
\bibliography{supporting_statement_arxiv}

\end{document}